\documentclass[aps,pra,twocolumn,amsmath,amssymb,showpacs]{revtex4}
\usepackage[T1]{fontenc}
\usepackage{amssymb,amsmath}
\usepackage{amsfonts}
\usepackage{graphicx}

\newcommand{\bJ}{{\bf J}}
\newcommand{\bn}{{\bf n}}

\newcommand{\tr}{{\rm tr}}

\def\vec#1{\mathbf{#1}}
\def\ket#1{|#1\rangle}
\def\bra#1{\langle#1|}
\def\ketbra#1{|#1\rangle\langle#1|}

\def\idmat{\mathbf{1}}
\def\jt{J_{\mathbf{t}}}
\def\ii{I\!I}

\input{tcilatex}

\begin{document}

\title{Classicality of spin states}
\author{Olivier Giraud$^{1}$, Petr Braun$^{2,3}$ and Daniel Braun$^{1}$}
\affiliation{$^{1}$ Laboratoire de Physique Th\'eorique,
  Universit\'e de Toulouse, CNRS,  31062 Toulouse, FRANCE\\
$^{2}$ Fachbereich Physik, Universit\"at Duisburg--Essen, 47048 Duisburg,
  GERMANY \\
  $^{3}$ Institute of Physics, Saint-Petersburg University, 198504 Saint-Petersburg, RUSSIA }
\begin{abstract}
We extend the concept of classicality in quantum optics to spin states. We
call a state ``classical'' if its density matrix can be
decomposed  as a weighted sum of angular momentum
coherent states with positive weights.
Classical spin states form a  convex
set $\cal C$, which we fully characterize for a spin--1/2
and a spin--1. For arbitrary spin, we provide
``non-classicality witnesses''.
For bipartite systems, $\cal C$ forms
a subset of all separable states. A state of two spins--1/2
belongs to $\cal C$ if and only if it is separable, whereas for a spin--1/2
coupled to a
spin--1, there are separable states which do not belong to $\cal C$.
We show that in general the question whether a state is in $\cal C$ can be
answered
 by a linear programming algorithm.
\end{abstract}
\pacs{02.40.Ft, 03.67.-a, 03.67.Mn}
\maketitle

\section{Introduction}

The question of the classicality of quantum states has regained
interest with the rise of quantum information theory
\cite{Nielsen00}. Stronger--than--classical correlations between
different systems are an important resource for quantum
communication protocols, and the existence of large amounts of
entanglement has been shown to be necessary for a quantum
computational speed-up \cite{Jozsa03,Vidal03}. However, even for a
single system the question of classicality is important.
Historically the question goes back to two seminal papers in
quantum optics by Sudarshan and Glauber
\cite{Sudarshan63,Glauber63}, who introduced the
Glauber--Sudarshan $P$--representation for the states of a harmonic
oscillator. This representation allows to decompose the density
matrix in terms of coherent states of the harmonic oscillator. For
a single coherent state, the weight function of the
$P$--representation (called $P$--function in the following for
short) reduces to a delta function on the phase space point in
which the coherent state is centered, and the dynamics of the
$P$--function is exactly the one of the classical phase space
distribution. It has therefore become customary in quantum optics
to consider states with a positive $P$--function as classical.
Several other criteria can be derived from this requirement. Using
Bochner's theorem for the Fourier transform of a classical
probability distribution \cite{Bochner33}, Richter and Vogel
derived a hierarchy of observable criteria based on the
characteristic function, which are both necessary and sufficient
for classicality \cite{Richter02}. This led to a recent
demonstration of the negativity of the $P$--function
in a quantum optical experiment \cite{Kiesel08}. Korbicz {\em et
  al.}~realized a connection of the
positivity of the $P$--function to Hilbert's 17th problem  of the
decomposition of a positive polynomial \cite{Korbicz05}. Since the
$P$--function for a continuous variable system can be highly
singular, a lot of attempts to define classicality have been based
on other quasi-probability distributions \cite{Cahill69} as well,
notably the Wigner function \cite{Hudson74,Kenfack04}.

These quasiprobability distributions for the harmonic oscillator
\cite{Cahill69} have analogs for finite--dimensional angular
momentum states \cite{Agarwal81}. The Wigner function for finite--dimensional
systems has received a large amount of attention, ranging from
questions of its most appropriate definition
\cite{Hannay80,Agarwal81,Dowling94,Rivas99,Bianucci02}, over
classicality criteria \cite{Cormick06,Gross06}, to the importance
of its negativity for quantum computational speed-up
\cite{Galvao05} (see also for further references concerning the
historical development of the Wigner function for finite--dimensional systems).
Surprisingly, the $P$--function for finite--dimensional systems has been much
 less studied, in spite of its
attractive mathematical properties. The $P$--function for a
system with a finite--dimensional Hilbert space (i.e.~formally a
spin system) allows to decompose the density matrix
%of a spin system
in terms of angular momentum coherent states
\cite{Arecchi72}. It can always be chosen to be a smooth function,
expandable in a finite set of spherical harmonic functions
\cite{Agarwal81}. In contrast to the case of the harmonic oscillator,
 questions
 concerning the existence of the
$P$--function (or its nature as a distribution or worse) do
therefore not arise. This idyllic situation is somewhat perturbed,
however,  by the fact, already observed in \cite{Arecchi72}, that
for a spin system a large amount of freedom exists in the choice
of the $P$--function, as it depends on two continuous variables on
the Bloch sphere, whereas the density matrix for a system with
$d$--dimensional Hilbert space is specified by $d^2-1$ real
independent entries.

In this paper we show that the existence of a $P$--representation
of the state of a spin system with a positive $P$--function is a
meaningful concept which allows to define the classicality of
states of finite--dimensional systems in a natural fashion,
completely analogous to the classicality of the harmonic
oscillator states of the electromagnetic field. We shall call the
corresponding states ``$P$--representable'', or $P$--rep for short.
The set $\cal C$ of $P$--representable states form a
convex domain in the space of density operators, containing the
completely mixed state in its interior. We show that,
surprisingly, all states of a single spin--1/2 are $P$--rep, and
obtain an analytical criterion for $P$--representability  in the
case of a spin--1. For bipartite systems, the set of $P$--rep states
is a subset of the set of separable states. For two spins--1/2 the
two sets coincide, whereas already for a spin--1/2 combined with a
spin--1, there are separable states which are not $P$--rep. We
also show that the problem of deciding whether a given state is
$P$--rep can be solved numerically by linear programming.

In the following we will first motivate and define $P$--representability,
then study simple cases of small spins,
introduce a variational approach that gives rise to a linear programming
algorithm, and finally have a look at composite systems. We also develop
some necessary conditions for $P$--representability based on measurable
observables, which may thus serve as ``non--classicality witnesses'', an
extension of the by now well-known concept of entanglement witnesses
\cite{Lewenstein00}.

\section{Definition of $P$--representability}

\subsection{Coherent states}
We first set some notations following the lines of \cite{Agarwal81}.
Angular momentum coherent states are defined as eigenstates of $\mathbf{J}^{2}$
and $\mathbf{n}.\mathbf{J}$ with eigenvalues $j(j+1)$ and $j$,
respectively,  where $\mathbf{n}$
 is a unit column vector which specifies the quantization axis with polar
 angle $\theta $ and
azimuth $\varphi$, and $\bJ$ is the familiar angular momentum operator with
components $J_x,J_y$ and $J_z$.  The transpose of the column vector $\bn$ reads
\begin{equation*}
\mathbf{n}\left( \theta ,\varphi \right)^t \mathbf{=}\left( \sin \theta \cos
\varphi ,\sin \theta \sin \varphi ,\cos \theta \right) .
\end{equation*}%
An angular momentum coherent state can be expanded in terms of the states
$\left\vert
jm\right\rangle $ quantized on the $z$ axis as
\begin{eqnarray*}
\left\vert \theta \varphi \right\rangle &=&\sum_{m=-j}^{j}
\sqrt{\binom{2j}{j+m}}\\
&\times&
 \left( \sin \frac{\theta }{2}\right) ^{j-m}\left( \cos \frac{\theta
}{2}\right) ^{j+m}e^{-i(j+m)\varphi }\left\vert jm\right\rangle .
\end{eqnarray*}
The coherent states form a complete, although not orthogonal, basis set of
normalized states
within the space of the eigenfunctions of $\mathbf{J}^{2}$ with given $j$, and
\begin{equation}
\frac{2j+1}{4\pi }\int \sin \theta d\theta d\phi \left\vert \theta \varphi
\right\rangle \left\langle \theta \varphi \right\vert =\idmat_{2j+1},\label{id}
\end{equation}
where $\idmat_{2j+1}$ is the $(2j+1)$--dimensional identity matrix.
We shall use the shorthand $\alpha =(\theta ,\varphi )$ and denote $d\alpha
=\sin \theta d\theta d\phi $. The coherent state $\left\vert \theta \varphi
\right\rangle$ associated with the
vector $\mathbf{n}$ will be denoted
$\left|\mathbf{n}\right\rangle$ or $|\alpha\rangle$.

\subsection{$P$--representation}
The $P$--representation of a density operator $\rho$ is an expansion over the
overcomplete basis of coherent states. This expansion
reads
\begin{equation}
\rho =\int d\alpha P\left( \alpha \right) \left\vert \alpha \right\rangle
\left\langle \alpha \right\vert,
\label{rhoP}
\end{equation}
where the $P$--function $P\left( \alpha \right)$ is real  and normalized by
the condition%
\begin{equation}
\tr\rho =\int d\alpha P\left( \alpha \right) =1\,.  \label{norma}
\end{equation}%
If $P\left( \alpha \right) $ is non-negative then $\rho$
is a classical mixture of pure coherent states with probability
density $P\left( \alpha \right) $, and can therefore be considered as
classical. In this case we shall say that $\rho $ is
$P$--representable, or \textquotedblleft $P$--rep\textquotedblright\ for short.

This definition has to be made more precise considering that $P(\alpha )$ is
not uniquely determined by the density operator. To show this, consider the
multipole expansion of $\rho $,
\begin{eqnarray}
\rho  &=&\dsum\limits_{K=0}^{2j}\dsum\limits_{Q=-K}^{K}\rho _{KQ}\widehat{T}%
_{KQ},\quad\ \  \rho _{KQ}=\tr\rho \widehat{T}_{KQ}^{\dagger },
\label{rhoinspherbas} \\
\hat{T}_{KQ}
&=&\sum_{m_{1,}m_{2}}^{j}(-1)^{j-m+Q}C_{jm_{1}jm_{2}}^{KQ}\left\vert
jm_{1}\right\rangle \left\langle jm_{2}\right\vert
\end{eqnarray}%
where $C_{jm_{1}jm_{2}}^{KQ}$ are the Clebsch-Gordan coefficients as
\cite{Abramowitz65}. Expanding 
the $P$--function as a sum of spherical harmonics,
\begin{equation*}
P(\alpha )=\dsum\limits_{K=0}^{\infty
}\dsum\limits_{Q=-K}^{K}P_{KQ}Y_{KQ}(\alpha ),
\end{equation*}
%it follows from (\ref{norma}) that $P_{00}=1/\sqrt{4\pi }.$There is a
one obtains a one-to-one relation between the coefficients of the two
expansions for $0\leq K\leq 2j$,
\begin{equation}
\rho _{KQ}=P_{KQ}\sqrt{4\pi }\frac{\left( 2j\right) !%
}{\sqrt{\Gamma(2j-K+1)\Gamma(2j+K+2)}}.  \label{ptorho}
\end{equation}%
If $K>2j$ the Euler Gamma functions in the denominator become infinite;
consequently
regardless of $P_{KQ}$ the respective $\rho _{KQ}$ will be zero. It means
that the choice of such $P_{KQ}$ is totally arbitrary. However,
non-negativity of a $P(\alpha)$ for one choice of $P_{KQ}$ with $K>2j$ may be
absent for another choice. Here is a simple example. Let the density
operator be a projector on a coherent state, $\rho =\left\vert \alpha
_{0}\right\rangle \left\langle \alpha _{0}\right\vert $. An obvious $P$%
-function in this case is $\delta (\alpha -\alpha _{0})$; it can
be considered non-negative since it can be approached by a
sequence of non-negative functions, like Gaussians with decreasing
width. An alternative choice however would be to drop all
non-physical terms in $P$ with $K>2j$, replacing the
$\delta$-function by a finite linear combination
\begin{equation*}
P(\alpha )=\dsum\limits_{K=0}^{2j}\dsum\limits_{Q=-K}^{K}Y_{KQ}^{\ast
}(\alpha _{0})Y_{KQ}(\alpha )
\end{equation*}%
which is \emph{not} non-negative for all finite $j$ (its tail away from the
maximum at $\alpha =\alpha _{0}$ oscillates around zero).

In view of the non-uniqueness of $P(\alpha )$ we reformulate the definition
of $P$--representability demanding that the condition $P\geq 0$ \ must be
fulfilled \emph{at least for one} particular $P(\alpha )$. Under this
definition the pure coherent state $\rho =\left\vert \alpha
_{0}\right\rangle \left\langle \alpha _{0}\right\vert $ will be $P$--rep,
which is intuitively reasonable. We are thus led to the following definition:
\begin{definition}
A density matrix $\rho$ is called $P$--rep
if it can be written as a convex sum of coherent states, i.e. as in
Eq.~\eqref{rhoP} with a non-negative function $P(\alpha)$.
\end{definition}
We will now derive some simple consequences of this definition.

\subsection{Consequences}
\label{consequences}
Let $\mathcal{V}$ be the vector space of $(2j+1)\times(2j+1)$
hermitian matrices. The scalar product $\langle X,Y\rangle=\tr
X^{\dagger}Y$ defines an operator norm $||X||=\sqrt{\tr
X^{\dagger}X}$ on $\mathcal{V}$. We denote by $\cal N$ the subset
of non-negative density matrices, and by $\cal C$ the subset of
$P$--rep states. The boundaries of these sets are respectively
denoted $\partial\cal N$ and $\partial\cal C$.
The following statements follow immediately from the above definition:

\begin{enumerate}
\item The totally mixed state
$\rho _{0}\equiv \frac{1}{2j+1}\idmat_{2j+1}$
is $P$--rep, which is readily seen from Eq.~(\ref{id}) taking $P\left(
\alpha \right) =1/4\pi$.

\item The set $\cal C$ of $P$--rep states is the convex hull of the set of
coherent states. In particular, it is a convex set.

\item Since all $P$--rep states are non-negative (but not vice versa)
we have $\mathcal{C}\subseteq\mathcal{N}\subseteq\mathcal{V}$.

%PB:
\item According to  Carath\'eodory's theorem on convex sets
applied to the $(2j+1)^2$--dimensional vector space $\mathcal{V}$,
any non-negative Hermitian matrix can be represented as a convex
sum of at most $(2j+1)^2+1$ projectors onto coherent states. In
the case of density matrices subject to the condition $\tr \rho=1$
this number is decreased by 1. Finding a $P$--representation for a
state $\rho$ is thus equivalent to finding real non-negative
coefficients $\lambda_i$ and coherent states $\ket{\alpha_i}$ such
that
   \begin{equation}
     \label{decompcs}
     \rho=\sum_{i=1}^{(2j+1)^2}\lambda_i\ket{\alpha_i}\bra{\alpha_i}.
   \end{equation}

\item A pure state is $P$--rep if and only if it is a coherent state. \\
{\em Proof.} The ``if'' part is trivial. For the ``only if'' part, assume
   that a state $\rho$ is $P$--rep, i.e. that there exists a
decomposition such as in \eqref{decompcs}. We have
   $\tr
   \rho^2=\sum_{i,j} \lambda_i\lambda_j |\langle
   \alpha_i|\alpha_j\rangle |^2\le \left(\sum_i \lambda_i\right)^2=1$, where
   equality occurs only for $|\langle\alpha_i|\alpha_j\rangle|=1$ for all
   $i,j$. The latter condition can only be fulfilled if there is  a single
   term in the sum. Thus a pure $P$--rep state, for which $\tr\rho^2=1$, has
   to be a coherent state.

\item Any density matrix can be decomposed as a sum of the totally mixed state $\rho_0$
and a traceless hermitian operator $\hat{\rho}$ with trace norm one  multiplied by a positive real
parameter $\kappa$,
\begin{equation}\label{hrho}
\rho_\kappa =\rho _{0}+\kappa\hat{\rho}.
\end{equation}
Since $\mathcal{C}$ is convex, there is, for any
given direction $\hat{\rho}$, an extremal value
$\kappa_{e}$ of $\kappa$ such that $\rho_\kappa\in\mathcal{C}$
if $0\leq \kappa < \kappa_{e}$ and
$\rho_\kappa\notin\mathcal{C}$ if
$\kappa>\kappa_{e}$.
The states $\rho=\rho_0+\kappa_{e}\hat{\rho}$
form the boundary $\partial\mathcal{C}$ of $P$--rep states.
They belong to $\mathcal{C}$ provided we
accept states $\rho$ as $P$--rep if they
can be approximated in the trace norm by a convex sum of
coherent states, that is for all $\epsilon>0$ there exists a positive
function $P(\alpha)$ such that
$\left\vert\left\vert\rho- \int d\alpha
P\left( \alpha \right) \left\vert \alpha
\right\rangle\left\langle \alpha \right\vert\right\vert\right\vert<\epsilon$.
With this extended definition the set of $P$--rep states becomes compact.
 In some directions the boundary
$\partial\mathcal{C}$ may touch $\partial\cal N$, e.g.~when $\rho
=\left\vert \alpha \right\rangle \left\langle \alpha \right\vert $ is
a pure coherent state.

\item $\partial\mathcal{C}$ is separated by a finite distance from the
 state $\rho_0$. In other words, all density operators in some finite
 neighborhood
 of $\rho_{0}$ are $P$--rep. To show it let us choose $P\left( \alpha \right)$
containing only the mandatory components with $K\leq 2j,$%
\begin{eqnarray}
\label{expansionPY}
P\left( \alpha \right)  &=&\frac{1}{4\pi }+\hat{P}\left( \alpha \right),\nonumber\\
\hat{P}\left( \alpha \right)
&=&\dsum\limits_{K=1}^{2j}\dsum\limits_{Q=-K}^{K}P_{KQ}Y_{KQ}(\alpha ).
\end{eqnarray}
The $P_{KQ}$ are bounded since they are related to the coordinates
$\rho_{KQ}$ of $\rho$ by (\ref{ptorho}) and $\tr\rho^2\leq 1$.
As the spherical harmonics are bounded on the sphere and
\eqref{expansionPY} is a finite sum, there is an upper bound $\hat{P}_{e}$
to the non-trivial part $\hat{P}\left(\alpha \right)$
when $\rho$ and $\alpha$ are varied. Thus, all matrices
$\rho _{0}+\kappa \hat{\rho}$ with $\kappa<1/(4\pi\hat{P}_{e})$
will be $P$--rep. 
\end{enumerate}

\section{$P$--rep for systems of small spin}

In the case of a spin--$1/2$ or a spin--1, it is possible to obtain a complete
 characterization of $P$--rep states.

\subsection{Spin--1/2}

We denote by $\mathbf{\sigma =}\left( \sigma _{x},\sigma _{y},\sigma _{z}\right) $
 the vector formed by the Pauli matrices. Together with the identity matrix
$\idmat_{2}$ they form a basis of the space of $2\times 2$ matrices.
Any $2\times 2$ Hermitian matrix with unit trace can be written as
\begin{equation}
\label{rho12}
\rho=\frac{1}{2}\left(\idmat_{2}+\mathbf{u}.\mathbf{\sigma}\right),
\end{equation}
and $\mathbf{u}$ is given by $\mathbf{u}=\tr(\rho\mathbf{\sigma})$.
The matrix $\rho $ is non-negative if and only if
$|\mathbf{u}|\leq 1$. A physical density matrix $\rho$ can thus
be represented by a point inside the unit sphere (the Bloch
sphere). Matrices corresponding to points on the unit sphere
 are pure states. Since for spin--1/2 any pure state is a coherent state,
the convex hull of coherent states is the convex hull of pure states, which
is the set of all density matrices. Thus all states are
$P$--rep.

It is straightforward  to find an explicit decomposition in terms
of angular momentum coherent states by simply diagonalizing
$\rho$, which leads to the sum of two projectors with two positive
eigenvalues. Nevertheless, there is a large freedom in choosing
the coherent states. According to \eqref{decompcs}, finding a
$P$--representation for $\rho$ amounts
 to finding positive real coefficients $\lambda_i$ and projectors on coherent states
$\ket{\alpha_i}\bra{\alpha_i}=\frac{1}{2}\left(\idmat_{2}+\mathbf{n}^{(i)}.\mathbf{\sigma}\right)$
with
 $|\mathbf{n}^{(i)}|=1$ such that $\rho=\sum_i\lambda_i
 \ket{\alpha_i}\bra{\alpha_i}$.
Since the $\sigma_i$ form a basis of the $2\times 2$ density matrices,
this is equivalent to finding $\lambda_i$ and norm-1 vectors $\mathbf{n}^{(i)}$
 such that
\begin{equation}
\mathbf{u}=\sum_i\lambda_i\mathbf{n}^{(i)}.
\end{equation}
This can be trivially achieved e.g.~by taking any pair of points on the Bloch sphere
such that the line joining these two points contains the point representing
$\mathbf{u}$ inside the sphere.

\subsection{Spin--1}
Let us now consider a spin--1 density matrix. We shall use the
representation
\begin{equation}
\rho =\frac{1}{3}\idmat_{3}+\frac{1}{2}\mathbf{u}.\mathbf{J}
+\frac{1}{2}\sum_{a,b=x,y,z}\left( W_{ab}-\frac{1}{3}\delta
_{ab}\right) \frac{J_{a}J_{b}+J_{b}J_{a}}{2}\,,
\label{canonrhoj1}
\end{equation}
where $J_{a}$ are matrices of the angular momentum with $j=1$. The $J_{a}$
and the $(J_{a}J_{b}+J_{b}J_{a})/2 $, together with the identity matrix
$\idmat_3$,
form a basis of the vector space $\mathcal{V}$ of $3\times 3$ hermitian
matrices.
Inverting relation (\ref{canonrhoj1}) we obtain
\begin{equation}
\label{uW}
u_{a}=\tr\left( \rho J_{a}\right) ,\quad W_{ab}=\func{Tr}\rho \left(
J_{a}J_{b}+J_{b}J_{a}\right) -\delta _{ab},
\end{equation}
which shows that $\mathbf{u\in }\mathbb{R}^{3}$ while $W$ is a
$3\times 3$ real symmetric tensor with
trace 1.
The projector on a coherent state $\ket{\mathbf{n}}$, written in the
form (\ref{canonrhoj1}), reads
\begin{equation}
\ketbra{\mathbf{n}}=\frac{1}{3}\idmat_{3}+\frac{1}{2}\mathbf{n}.\mathbf{J}
+\frac{1}{2}\sum_{a,b=x,y,z}\left( n_{a}n_{b}-\frac{1}{3}\delta _{ab}\right) \frac{%
J_{a}J_{b}+J_{b}J_{a}}{2}.
\label{cohj1}
\end{equation}
According to \eqref{decompcs}, $\rho$ is $P$--rep if and only if there exist
$\lambda_i>0$ with $\sum_i\lambda_i=1$ and coherent states corresponding to vectors
$\mathbf{n}^{(i)}\in\mathbb{R}^{3}$ of length 1 such that
\begin{eqnarray}
\label{equations_spin1}
\sum_{i}\lambda _{i}n^{\left( i\right) }_a&=&u_a,  \label{eqlamnj1} \\
\sum_{i}\lambda _{i}n^{\left( i\right) }_a n^{\left( i\right) }_b&=&W_{ab},\notag
\end{eqnarray}
(with $a,b$ running over $x,y,z$). It turns out that these equations
admit a solution -- and hence $\rho$ is $P$--rep -- if and only if
the real symmetric $3\times 3$ matrix $Z$ with matrix elements
\begin{equation}
\label{defZ}
Z_{ab}=W_{ab}-u_a u_b
\end{equation}
is non-negative.\\
{\em Proof.} First let us assume that the Eqs.~(\ref{eqlamnj1})
do have a solution.
Then $Z$ can be written
\begin{equation}
Z_{ab}=\sum_{i,j}\left( \lambda _{i}\delta _{ij}-\lambda _{i}\lambda _{j}\right)
n^{(i)}_a n^{(j)}_b,
\end{equation}
and for any vector $\mathbf{y}$\ $\in $ $\mathbb{R}^{3}$ we have
\begin{equation}
\mathbf{y}^tZ\mathbf{y}
=\sum_{i}\lambda _{i}\left(\mathbf{y}.\mathbf{n}^{\left( i\right)}\right)^{2}
-\left( \sum_{i}\lambda _{i}\mathbf{y}.\mathbf{n}^{\left( i\right) }\right)^{2}
\geq 0
\label{yZyj1}
\end{equation}
since the weights $\lambda _{i}>0$ sum to 1 and $f(x)=x^{2}$ is a convex
function. Therefore $Z$ is indeed non-negative for all $P$--rep operators $\rho $.

Conversely, if $Z\geq 0$, then it is possible to exhibit a decomposition
of $\rho$ by finding an explicit solution to Eqs.~\eqref{equations_spin1}.
Let $A$ be such that $Z=A{}A^t$.
If we denote by $\mathbf{t}^{(i)}$ the eight column
vectors $(\pm 1,\pm 1,\pm 1)$ obtained from all combinations of the $\pm$
signs, and define
\begin{equation}
\tau_i=-\frac{\mathbf{u}^tA\mathbf{t}^{(i)}}{1-|\mathbf{u}|^2}
+\sqrt{1+\left(\frac{\mathbf{u}^t A\mathbf{t}^{(i)}}{1-|\mathbf{u}|^2}\right)^2},
\end{equation}
then one can check that a solution to Eqs.~\eqref{equations_spin1} is given by
\begin{eqnarray}
\mathbf{n}^{(i)}&=&\mathbf{u}+\tau_iA\mathbf{t}^{(i)}\\
\lambda_i&=&\frac{1}{4}\frac{1}{1+\tau_i^2}\,,
\end{eqnarray}
which proves that $\rho$ is $P$--rep.

The necessary and sufficient condition $Z\geq 0$ in the case of
spin--1 allows to characterize the boundary $\partial\mathcal{C}$
of $P$--rep states. Indeed, let us consider a one-parameter family
of states as in (\ref{hrho}). If $\mathbf{u}$ and $W$ are the
vector and matrix corresponding to the expansion \eqref{canonrhoj1} of the state
$\rho_0+\hat{\rho}$, then the vector and
the matrix associated with $\rho_{\kappa}=\rho_0+\kappa\hat\rho$ are given by
\begin{eqnarray}
\mathbf{u}_{\kappa }&=&\kappa \mathbf{u}\\
W_{\kappa }&=&\kappa W+\left( \frac{1-\kappa }{3}\right)
\idmat_{3}\nonumber,
\end{eqnarray}
and thus the $3\times 3$
matrix $Z_{\kappa}$ associated with  $\rho_{\kappa}$ reads
\begin{equation}
Z_{\kappa }=\kappa W+\left( \frac{1-\kappa }{3}\right)
\idmat_{3}-\kappa ^{2}\mathbf{u}\mathbf{u}^t.
\end{equation}
The value $\kappa=\kappa _{e }$ at which the scaled operator
$\rho _{\kappa}$ ceases to be $P$--rep corresponds to the smallest
$\kappa$ for which $Z_{\kappa}$ has a zero eigenvalue. Thus $\kappa_{e }$
is the smallest solution of the equation $\det Z_{\kappa}=0$, and the
equation of $\partial\mathcal{C}$ in the vector space $\mathcal{V}$ is
\begin{equation}
\kappa _{e }^{2}\mathbf{u}^t\left( \kappa _{e }W+
\frac{1-\kappa _{e }}{3}\idmat_{3}\right) ^{-1}\mathbf{u}=1.\label{ke}
\end{equation}
This equation  gives implicitly the value $\kappa_{e}$ for each direction
$\hat{\rho}$  in
the vector space $\mathcal{V}$. As the examples of spin--1/2 and spin--1
show, the proportion of $P$--rep matrices among 
all density operators depends on $j$.

\subsection{Necessary conditions for higher spins}
It is  possible to derive more general necessary conditions
for $P$--representability of spin--$j$ states, as follows. Let us denote
by $\jt=\mathbf{t.J}$ the spin operator in direction $\mathbf{t}$.
For a coherent state $\ket{\mathbf{n}}$ corresponding to a vector $\mathbf{n}$,
the mean values of $\jt$ and $\jt^2$ are given by
\begin{eqnarray}
\bra{\mathbf{n}}\jt\ket{\mathbf{n}}&=&j\ \mathbf{t}.\mathbf{n}\\
\bra{\mathbf{n}}\jt^2\ket{\mathbf{n}}&=&\frac{j}{2}
+j\left(j-\frac{1}{2}\right)\left(\mathbf{t}.\mathbf{n}\right)^2.
\end{eqnarray}
Any $P$--rep state $\rho$ can be written as
$\rho=\sum_i\lambda_i\ket{\mathbf{n}^{(i)}}\bra{\mathbf{n}^{(i)}}$, which
implies for the mean values of $\jt$ and  $\jt^2$ in the state $\rho$
\begin{eqnarray}
\langle\jt\rangle&=&j\ \sum_i\lambda_i\ \mathbf{t}.\mathbf{n}^{(i)}
\label{jt}\\
\langle\jt^2\rangle&=&\frac{j}{2}+j\left(j-\frac{1}{2}\right)
\sum_i\lambda_i\left(\mathbf{t}.\mathbf{n}^{(i)}\right)^2.
\label{jt2}
\end{eqnarray}
Convexity of $f(x)= x^{2}$ applied to the sums over $i$ leads to the
inequality
\begin{equation}
\label{ineq}
2j\langle\jt^2\rangle-(2j-1)\langle\jt\rangle^2-j^2\geq 0\ \ \ \ \forall
\mathbf{t},\ |\mathbf{t}|=1,
\end{equation}
with equality if and only if $\rho$ is itself a coherent state.
This is a necessary condition for $P$--rep, valid for any $j$.
In the particular case of spin--1/2 this inequality becomes
$\langle\jt^2\rangle\geq 1/4$, which is obviously true for all
states $\rho$ and all directions $\mathbf{t}$. In the case of spin--1
the inequality \eqref{ineq} can be rewritten as
\begin{equation}
\sum_{a,b}\left(2\langle J_a J_b\rangle-\langle J_a\rangle\langle J_b\rangle
-\delta_{ab}\right)t_a t_b\geq 0\ \ \ \ 
\forall \mathbf{t}=(t_x,t_y,t_z),\ |\mathbf{t}|=1.
\end{equation}
As can be  seen from Eqs.~\eqref{uW} and \eqref{defZ}, 
this inequality exactly corresponds to the condition $Z\ge 0$ derived
 in the previous section.

For higher spins, one can similarly derive other
necessary conditions. For instance
for a $P$--rep state of spin--3/2, one has
\begin{equation}
\langle\jt^3\rangle=\frac{21}{8}
\sum_i\lambda_i\left(\mathbf{t}.\mathbf{n}^{(i)}\right)+\frac{3}{4}
\sum_i\lambda_i\left(\mathbf{t}.\mathbf{n}^{(i)}\right)^3,
\end{equation}
and a necessary condition imposed by the fact that
$|\sum_i\lambda_i x_i^3|\leq\sum_i\lambda_i x_i^2$ for any $x_i\in[-1,1]$
reads
\begin{equation}
\forall
\mathbf{t},\ \ \ 2\left|\langle\jt^3\rangle-\frac{7}{4}\langle\jt\rangle\right|
\leq\left|\langle\jt^2\rangle-\frac{3}{4}\right|.
\end{equation}
These necessary conditions can be considered as ``non-classicality
witnesses'', as a state $\rho$ is not in $\mathcal{C}$ if at least one of these
conditions is not fulfilled.

\section{Numerical implementation}

\subsection{Variational approach to $P$--representability}

Suppose we are given a density operator and want to establish whether it is $%
P$--representable. Let us use the multipole expansion (\ref{rhoinspherbas}).
The coefficients $P_{KQ}$ with $0\leq K\leq 2j$ will be defined by
Eq.~\eqref{ptorho}. Orthogonality of the spherical harmonics implies that
the hypothetical $P(\alpha)\geq 0$ 
satisfies the integral equations
\begin{equation}
\int P(\alpha )Y_{KQ}^{\ast }(\alpha )d\alpha =P_{KQ},\quad 0<K\leq 2j,\quad
|Q|\leq K,  \label{kbigger0}
\end{equation}
together with
\begin{equation*}
\text{ }\int P(\alpha )d\alpha =\tr\rho =1.
\end{equation*}
If we find any $P(\alpha )\geq 0$ satisfying these equations the state in
question is $P$--representable.

We can ask for more and try to find the representability boundary
for all matrices of the form $\rho _{\kappa }=\rho_0+\kappa\hat{\rho}$
obtained by scaling a given traceless normalized hermitian matrix
$\hat{\rho}$. To that end, we consider the set of matrices
$\rho_0/\kappa+\hat{\rho}$, $\kappa>0$. These states all have the same
traceless part $\hat\rho$, thus they are
represented by $P$--functions $P(\alpha)$ that satisfy
Eqs.~\eqref{kbigger0} with $P_{KQ}$ corresponding to $\hat{\rho}$,
but with $\int P(\alpha )d\alpha=\frac{1}{\kappa}$. We look at the
minimum of the functional $F\left[ P\right] \equiv \int P(\alpha
)d\alpha$ over these states. Suppose that the minimum is realized
by some function $P_{e}(\alpha )$ and introduce $\kappa _{e} $
through
\begin{equation}
\min \int P(\alpha )d\alpha =\int P_{e}(\alpha )d\alpha =\frac{1}{\kappa _{e}%
}.  \label{varpin}
\end{equation}%
The corresponding density operator
$\rho_{\kappa _{e}}=\rho_{0}+\kappa _{e}\hat{\rho}$
is represented by the function $\kappa _{e}P_{e}(\alpha )$. As we pointed
out it means that all operators $\rho_\kappa$ with $0\leq\kappa <\kappa _{e}$
are $P$--representable and that $\rho_e$ belongs to the boundary
$\partial\mathcal{C}$.

\subsection{Concavity of $1/\kappa_{e}$}
The parameter $\kappa _{e}$ corresponding to the border of $P$--rep depends
on the matrix $\rho ,$ such that $\kappa _{e}=\kappa _{e}(\rho )$. Let us
take two matrices, $\rho ^{I}$ and $\rho ^{\ii}$ and calculate the respective
$\kappa _{e}(\rho ^{I})$, $\kappa _{e}(\rho ^{\ii})$. Consider now a convex
combination
\begin{equation*}
\rho ^{(c)}=c\rho ^{I}+(1-c)\rho ^{\ii},\quad 0<c<1.\quad
\end{equation*}
Then
\begin{equation*}
\frac{1}{\kappa _{e}(\rho ^{(c)})}\leq \frac{c}{\kappa _{e}(\rho ^{I})}+
\frac{1-c}{\kappa _{e}(\rho ^{\ii})},
\end{equation*}
i.e., $1/\kappa _{e}$ is a concave function of $\rho $.
The proof is based on Eq.~(\ref{varpin}). Let $P_{e}^{I},P_{e}^{\ii}$ be the
functions minimizing $\int Pd\alpha $ under constraints corresponding to the
operators $\rho ^{I}$ and $\rho ^{\ii}$ respectively. Then the function $%
P^{(c)} =cP_{e}^{I}+(1-c)P_{e}^{\ii}$ will obey the
constraints corresponding to the operator $\rho^{(c)}$. Therefore we must
have
\begin{eqnarray*}
\frac{1}{\kappa _{e}(\rho ^{(c)})} &=&\min \int P(\alpha)\,d\alpha \leq \int P^{(c)}(\alpha)
\,d\alpha \\
&=&c\int P_{e}^{I}(\alpha)\,d\alpha +(1-c)\int P_{e}^{\ii}(\alpha)\,d\alpha \\
&=&\frac{c}{\kappa_{e}(\rho ^{I})}+\frac{1-c}{\kappa _{e}(\rho ^{\ii})},
\end{eqnarray*}
which implies concavity of $1/\kappa_e$. Thus the knowledge of $\kappa_e$
for two density matrices gives a lower bound for
a whole family of convex combinations of these density matrices.

\subsection{Linear programming}
In order to numerically implement the variational approach described here, 
let us choose the trial $P$--function in the form of a linear
combination of $\delta$-peaks
\begin{equation}\label{linkco}
P\left(\alpha\right)=\sum_{i=1}^n
w_i\delta\left(\alpha-\alpha_i\right)
\end{equation}
where the points $\alpha_i=(\theta_i,\phi_i)$ are more or less
uniformly distributed on the unit sphere, and $w_i\ge 0$ are
non-negative variational parameters; the delta-functions are
assumed to be normalized on the unit sphere,
$\delta\left(\alpha-\alpha_i\right)=\delta\left(\cos \theta-\cos
\theta_i\right)\delta\left(\phi-\phi_i\right)$. Inserting this $P(\alpha)$
in (\ref{kbigger0}) we come to the optimization problem: find
$\vec w=\{w_1,\ldots,w_n\}$ with all $w_i\ge 0,\;i=1\ldots n,$
minimizing the sum
\begin{equation}
F(\vec w)=\sum_{i=1}^n w_i,
\end{equation}
and subject to $M=(2j+1)^{2}-1$ linear constraints%
\begin{equation*}
\sum_{i=1}^{n}Y_{KQ}(\alpha _{i})\; w_{i}=P_{KQ},\quad 0<K\leq
2j,\quad |Q|\leq K.
\end{equation*}%
 This is a problem of linear programming \cite{Cormen01}. Its
well-known theorem states that whatever the number of unknowns $n$
the minimum of $F$ is realized on a solution  containing
no more than $M$ non-zero components. This number is one less than
predicted by Caratheodory's theorem because the solution  is a
%PB
boundary, not an internal, point of the set of the density
matrices $P$-representable  by  (\ref{linkco}).
 The minimum found numerically for a given $n$ yields an upper bound
on the exact value of $1/\kappa _{e}$ (Eq.~(\ref{varpin})), i.e., the lower bound on the
value of the scaling parameter $\kappa$ at the border of $P$--rep
in $\rho _{\kappa }=\rho _{0}+\kappa \hat{\rho}$.

The linear programming approach was numerically tested and found efficient
for moderate values of $j$. For a given $\rho ,$ the minimal value of $%
\kappa ^{-1}$ diminished fast with the increase of $n$ and was
stable. On the other hand, the solution $\vec{w}$ changed
erratically with the change
of $n$. That was to be expected considering the freedom in the choice of
$P\left( \alpha \right)$.
\vspace{0.5cm}

\section{Composite systems}
The definition of classicality can be extended to systems of more than one
particle in a natural way. In the present section we shall consider the case
of two particles, but the formalism generalizes to an arbitrary
number of particles.

\subsection{Classicality for two particles}
The $P$--representation of a density operator in the case of two spins $j_{A}$%
and $j_{B}$,
\begin{equation}
\rho =\int d^{2}\alpha _{A}d^{2}\alpha _{B}P(\alpha _{A},\alpha
_{B})\left\vert \alpha _{A}\right\rangle \left\vert \alpha _{B}\right\rangle
\left\langle \alpha _{A}\right\vert \left\langle \alpha _{B}\right\vert
\end{equation}%
with $P\geq 0$ is possible for separable states only; consequently $P$--rep
is a sufficient criterion of separability. The partially transposed matrices $%
\rho^{T_A}$ and $\rho^{T_B}$ are defined in a fixed computational basis
$|ij\rangle\equiv|i\rangle_A\otimes|j\rangle_B$ as
$\rho_{ij,kl}^{T_A}=\rho_{kj,il}$ and
$\rho_{ij,kl}^{T_B}=\rho_{il,kj}$. They are $P$--rep if and only if $\rho $
is $P$--rep, and the corresponding $P$--functions $P^{T_A}$ and $P^{T_B}$ are
simply related to the $P$--function of $\rho$ by
$%
P^{T_A}(\alpha _{A},\alpha _{B})=P(\tilde{\alpha}_{A},\alpha _{B}),\quad
\tilde{\alpha}_{A}=\left(\theta _{A},-\varphi_{A}\right)$, and
correspondingly for $P^{T_B}$.
All previously considered equations are reformulated for two spins in a
straightforward manner; we shall list them without commenting.

The representation of $\rho $ in terms of products of spherical multipole
operators reads
\begin{widetext}
\begin{equation}
\rho
=\dsum\limits_{K_{A}=0}^{2j_{A}}\dsum\limits_{Q_{A}=-K_{A}}^{K_{A}}\dsum%
\limits_{K_{B}=0}^{2j_{B}}\dsum\limits_{Q_{B}=-K_{B}}^{K_{B}}\rho
_{K_{A}Q_{A,}K_{B}Q_{B}}\widehat{T}_{K_{A}Q_{A}}^{A}\widehat{T}%
_{K_{B}Q_{B}}^{B},
\end{equation}
and we have
the $P$--function expanded into products of spherical harmonics,%
\begin{equation*}
P(\alpha )=\dsum\limits_{K_{A}=0}^{\infty
}\dsum\limits_{Q_{A}=-K_{A}}^{K_{A}}\dsum\limits_{K_{B}=0}^{\infty
}\dsum\limits_{Q_{B}=-K_{B}}^{K_{B}}P_{K_{A}Q_{A,}K_{B}Q_{B}}Y_{K_{A}Q_{A}}(%
\alpha _{A})Y_{K_{B}Q_{B}}(\alpha _{B}).
\end{equation*}
The relation between the coefficients of $\rho $ and $P$ is given by%
\begin{eqnarray*}
\rho _{K_{A}Q_{A,}K_{B}Q_{B}} &=&P_{K_{A}Q_{A,}K_{B}Q_{B}} \\
&&\times 4\pi \frac{\left( 2j_{A}\right) !\left( 2j_{B}\right) !}{\sqrt{%
(2j_{A}-K_{A})!(2j_{A}+K_{A}+1)!(2j_{B}-K_{B})!(2j_{B}+K_{B}+1)!}},
\end{eqnarray*}
and the density operator with a scaled non-trivial part by%
\begin{eqnarray*}
\rho _{\kappa } &=&\rho _{0}+\kappa \hat{\rho}, \\
\rho _{0} &=&\frac{\idmat_{\left( 2j_{A}+1\right) \times \left(
2j_{B}+1\right) }}{(2j_{A}+1)(2j_{B}+1)}.
\end{eqnarray*}

The following variational problem needs to be solved when the boundary of $P$%
--representability is to be found: minimize the functional%
\begin{equation*}
F\left[ P\right] =\int d^{2}\alpha _{A}d^{2}\alpha _{B}P(\alpha _{A},\alpha
_{B})
\end{equation*}%
with $P(\alpha _{A},\alpha _{B})\geq 0$ satisfying the integral equations
\begin{equation}
\int d^{2}\alpha _{A}d^{2}\alpha _{B}P(\alpha _{A},\alpha
_{B})Y_{K_{A}Q_{A}}^{\ast }(\alpha _{A})Y_{K_{B}Q_{B}}^{\ast }(\alpha
_{B})=P_{K_{A}Q_{A},K_{B}Q_{B}}\,,
\end{equation}%
where $K_{A},K_{B}$ run from $0$ to $2j$ excluding $K_{A}=K_{B}=0$, and $|Q_{A}|%
\leq K_{A},|Q_{B}|\leq K_{B}$. If the minimum of $F$ is equal to%
\begin{equation*}
F_{e}=\min F=\int d^{2}\alpha _{A}d^{2}\alpha _{B}P_{e}(\alpha _{A},\alpha
_{B})\equiv \frac{1}{\kappa _{e}}\,,
\end{equation*}%
then the density operator lying on the boundary of $P$--representability will be
$\rho _{\kappa _{e}}$. 

For the numerical implementation, the integrals are now taken over
a product of two unit spheres of Alice and Bob. Let us choose the
trial $P$--function as
\begin{equation}
P\left(\alpha_A,\alpha_B\right)=\sum_{i_{A}=1}^{n_{A}}\sum_{i_{B}=1}^{n_{B}}w_{i_Ai_B}
\delta\left(\alpha_A -\alpha^A_{i_A}\right)\delta\left(\alpha_B
-\alpha^B_{i_B}\right)
\end{equation}
where $n_A$ points $\alpha^A_{i_A}$ and $n_B$ points
$\alpha^B_{i_B}$ are uniformly scattered over the spheres of Alice
and Bob, respectively, and $w_{i_Ai_B}\ge 0$ are $n_An_B$
variational parameters. We now solve the linear programming task:
minimize
\begin{equation*}
F\left( \vec{w}\right)
=\sum_{i_{A}=1}^{n_{A}}\sum_{i_{B}=1}^{n_{B}}w_{i_{A}\,i_{B}}
\end{equation*}%
with $w_{i_{A}\,i_{B}}\geq 0$ satisfying
$M=(2j_{1}+1)^{2}(2j_{2}+1)^{2}-1$
linear constraints,%
\begin{equation*}
\sum_{i_{A}=1}^{n_{A}}\sum_{i_{B}=1}^{n_{B}}Y_{K_{A}Q_{A}}^{\ast
}(\alpha _{i_{A}}^{A})Y_{K_{B}Q_{B}}^{\ast }(\alpha
_{i_{B}}^{B})\;w_{i_{A}i_{B}} =P_{K_{A}Q_{A},K_{B}Q_{B}} .
\end{equation*}
Here $K_{A},Q_{A},K_{B},Q_{B}$ take all possible values excluding $%
K_{A}=K_{B}=0 $. Again, the optimal solution contains no more than
$M$ non-zero elements $w_{i_{A}i_{B}}$.

\subsection{Two spins 1/2}

Considering that the density operator of a single spin--1/2 is always $P$--rep
it is easy to see that the density operator for a system of two spins is $P$-
rep if and only if it is separable. Consequently, the necessary and
sufficient condition of $P$--rep is given by the Peres-Horodecki theorem
\cite{Peres93, Horodecki96}. It means that the boundary of $P$--representability in the
family $\rho _{\kappa }=\rho _{0}+\kappa \hat{\rho}$ is reached when either
$\rho _{\kappa }$ or its partial transpose $\rho _{\kappa }^{T_A}$ ceases to
be non-negative. This was checked numerically in the linear programming
approach: the minima $1/\kappa _{e}$ of the functional $F\left[ P\right] $
calculated with the matrix $\rho $ and its partial transpose $\rho ^{T_A}$
in all cases coincided with each other and agreed with the scaling necessary
to shift the smallest eigenvalue of either $\rho $ or $\rho ^{T_A}$ to
zero. The optimal $P$ was obtained as a combination of $M=15$ coherent states,
some of them with very small weights.

\subsection{Spins 1/2 and 1}

In this case the separability and $P$--rep conditions do not coincide.
Indeed consider for instance the pure product state (in $|j m\rangle$
notation)
$\ket{\psi}=\ket{\frac{1}{2} \frac{1}{2}}\otimes\ket{1 0}$.
Then the mean value of the operator $\idmat_2\otimes J_z^2$ in the
state $\ket{\psi}$ is $\bra{1 0}J_z^2\ket{1 0}=0$, while using
Eq.~\eqref{jt2} one should have for a $P$--rep state
$\langle\idmat_2\otimes J_z^2\rangle\ge 1/2$. Thus, $\ket{\psi}$ is not
$P$--rep. More generally, it is easy to show numerically that $\partial
\cal{C}$ is well inside the separability boundary. An example is shown in
Fig.\ref{fig.2D}, where we display the two boundaries for a density matrix
of the form $\rho=\rho_0+\kappa_1\hat\rho_1+\kappa_2\hat\rho_2$ with two
random but fixed traceless parts $\hat{\rho}_1$ and $\hat{\rho}_2$.

\begin{figure}
\begin{center}
  \includegraphics[scale=0.6]{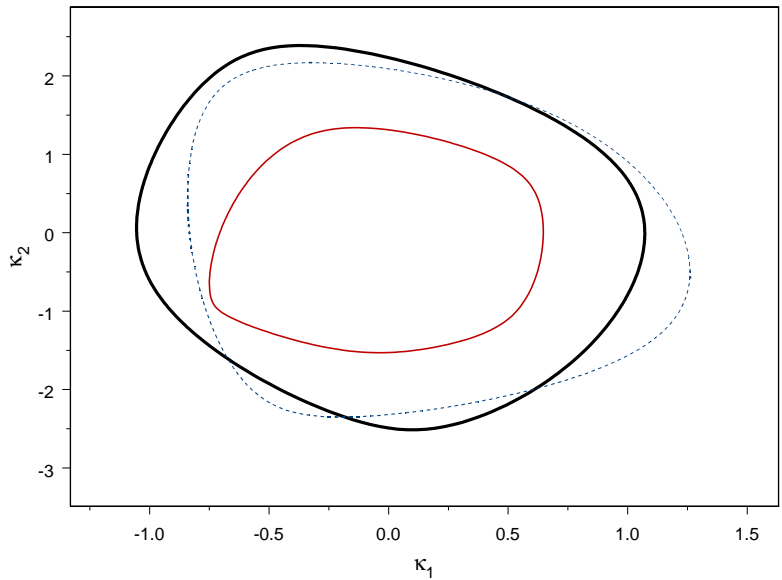}
\end{center}
\caption{(Color online) Example of a set of classical states
$\mathcal{C}$ for a bipartite system of two spins 1/2 and 1
parametrized by two parameters,
$\rho=\rho_0+\kappa_1\hat\rho_1+\kappa_2\hat\rho_2$ with some
traceless $\hat\rho_1,\;\hat\rho_2$. Boundaries are shown of
non-negativity of $\rho$ (bold black line), non-negativity of its
partial transpose $\rho^{T_A}$ (dashed line), and of
$P$--representability of $\rho,\;\rho^{T_A}$ (inner red line).}
\label{fig.2D}
\end{figure}

\subsection{Classicality witness}
A simple necessary condition for $P$--rep can be formulated for the density
operator $\rho $ of the system of two particles $A$ and $B$. Let $V_{A}$ be
any
non-negative operator in the Hilbert space of $A$
and take the partial trace of $\rho V_{A}$ over the $A-$variables. Assuming that
$\rho $ is $P$--rep and using the coherent states $\left\vert \alpha'
\right\rangle $ for the calculation of the trace we obtain%
\begin{eqnarray}
\func{Tr}_{A}\rho V_{A} &=&\frac{2j+1}{4\pi }\int d\alpha ^{\prime
}\left\langle \alpha ^{\prime }|\rho V_{A}|\alpha ^{\prime }\right\rangle  \\
&=&\frac{2j+1}{4\pi }\int d\beta \left\vert \beta \right\rangle \left\langle
\beta \right\vert \int d\alpha P\left( \alpha ,\beta \right) \int d\alpha
^{\prime }\left\langle \alpha |V_{A}|\alpha ^{\prime }\right\rangle
\left\langle \alpha ^{\prime }|\alpha \right\rangle  \\
&=&\int d\beta \bar{P}\left( \beta \right) \left\vert \beta \right\rangle
\left\langle \beta \right\vert
\end{eqnarray}
where $\bar{P}\left( \beta \right) =\int d\alpha P\left( \alpha ,\beta
\right) \left\langle \alpha |V_{A}|\alpha \right\rangle $ is manifestly
non-negative. Consequently,
\begin{equation}
\rho _{B}=\left( \func{Tr}_{A}\rho V_{A}\right) /\func{Tr}\rho V_{A}
\end{equation}
can be considered as a density operator in the $B-$space which is
$P$--representable
by a function $\bar{P}\left( \beta \right) /\func{Tr}\rho V_{A}$. Therefore
$\rho $ can be $P$--rep only if $\rho _{B}$ is also $P$--rep (not vice versa).
The $P$--rep of $\rho _{B}$ is easy to check using our result for $j=1$. One
can take, e.g., $V_{A}=\idmat_{A}$ getting $\rho _{B}=\func{Tr}_{A}\rho $.
\end{widetext}

\section{Conclusion}
The $P$--representable states are classical mixtures of projectors on
angular momentum coherent states, i.e.~of angular momentum states with
minimal uncertainty. The $P$--rep states have many 
interesting properties. They can be seen as the ``most classical'' states, an
\textquotedblleft inner 
circle\textquotedblright\  within the linear
space of density operators which forms a convex set $\mathcal{C}$ that contains
the totally mixed state in its interior.  In the case of two spins,
$\mathcal{C}$
is a subset of the set of separable states.
The study of the $P$--representation provides thus important information on the
structure of  space of density matrices.

We have studied conditions for $P$--representability, and
completely characterized the set of classical states  for small
spins: for a spin--1/2 all states are $P$--rep, and for a spin--1 we
deduced a necessary and sufficient condition for $P$--rep. In the
case of two spins--1/2, $P$--rep is equivalent to separability, but
already for a spin--1/2 combined with a spin--1, there are states
which are separable but not $P$--rep. In addition, we have shown
that the question whether a given state is $P$--rep or not can be
solved  with a practical numerical method based on the linear
programming algorithm for finding the border of $P$--rep. We have
also formulated necessary conditions based on measurable
observables for $P$--rep, which can be considered
``non-classicality witnesses'' for spin systems.

Both analytical and computational methods have been used so far on very
modest values of $j$ (up to $j\sim 2$); for large $j$ the numerical methods
become forbiddingly slow. It would be important to investigate the limit of
large $j$ and provide thus a bridge to the case of continuous variables
where the $P$--rep states were an object of intense studies for many years
and proved to be of great physical importance.

\begin{acknowledgments}  We thank IDRIS in Orsay and CALMIP in Toulouse
for use of their computers.
This work was supported in part by the Agence
National de la Recherche (ANR), project INFOSYSQQ, contract number
ANR-05-JCJC-0072, the EC IST-FET project EUROSQIP, and the SFB/TR12
of the DFG.
\end{acknowledgments}

\bibliography{../mybibs_bt}

\begin{thebibliography}{26}
\expandafter\ifx\csname natexlab\endcsname\relax\def\natexlab#1{#1}\fi
\expandafter\ifx\csname bibnamefont\endcsname\relax
  \def\bibnamefont#1{#1}\fi
\expandafter\ifx\csname bibfnamefont\endcsname\relax
  \def\bibfnamefont#1{#1}\fi
\expandafter\ifx\csname citenamefont\endcsname\relax
  \def\citenamefont#1{#1}\fi
\expandafter\ifx\csname url\endcsname\relax
  \def\url#1{\texttt{#1}}\fi
\expandafter\ifx\csname urlprefix\endcsname\relax\def\urlprefix{URL }\fi
\providecommand{\bibinfo}[2]{#2}
\providecommand{\eprint}[2][]{\url{#2}}

\bibitem[{\citenamefont{Nielsen and Chuang}(2000)}]{Nielsen00}
\bibinfo{author}{\bibfnamefont{M.~A.} \bibnamefont{Nielsen}} \bibnamefont{and}
  \bibinfo{author}{\bibfnamefont{I.~L.} \bibnamefont{Chuang}},
  \emph{\bibinfo{title}{Quantum Computation and Quantum Information}}
  (\bibinfo{publisher}{Cambridge University Press}, \bibinfo{year}{2000}).

\bibitem[{\citenamefont{Jozsa and Linden}(2003)}]{Jozsa03}
\bibinfo{author}{\bibfnamefont{R.}~\bibnamefont{Jozsa}} \bibnamefont{and}
  \bibinfo{author}{\bibfnamefont{N.}~\bibnamefont{Linden}},
  \bibinfo{journal}{Proc. R. Soc. Lond. A} \textbf{\bibinfo{volume}{459}},
  \bibinfo{pages}{2011} (\bibinfo{year}{2003}).

\bibitem[{\citenamefont{Vidal}(2003)}]{Vidal03}
\bibinfo{author}{\bibfnamefont{G.}~\bibnamefont{Vidal}},
  \bibinfo{journal}{Phys. Rev. Lett.} \textbf{\bibinfo{volume}{91}},
  \bibinfo{pages}{147902} (\bibinfo{year}{2003}).

\bibitem[{\citenamefont{Sudarshan}(1963)}]{Sudarshan63}
\bibinfo{author}{\bibfnamefont{E.~C.~G.} \bibnamefont{Sudarshan}},
  \bibinfo{journal}{Phys. Rev. Lett.} \textbf{\bibinfo{volume}{10}},
  \bibinfo{pages}{277} (\bibinfo{year}{1963}).

\bibitem[{\citenamefont{Glauber}(1963)}]{Glauber63}
\bibinfo{author}{\bibfnamefont{R.~J.} \bibnamefont{Glauber}},
  \bibinfo{journal}{Phys. Rev.} \textbf{\bibinfo{volume}{131}},
  \bibinfo{pages}{2766} (\bibinfo{year}{1963}).

\bibitem[{\citenamefont{Bochner}(1933)}]{Bochner33}
\bibinfo{author}{\bibfnamefont{S.}~\bibnamefont{Bochner}},
  \bibinfo{journal}{Math. Ann.} \textbf{\bibinfo{volume}{108}},
  \bibinfo{pages}{378} (\bibinfo{year}{1933}).

\bibitem[{\citenamefont{Richter and Vogel}(2002)}]{Richter02}
\bibinfo{author}{\bibfnamefont{T.}~\bibnamefont{Richter}} \bibnamefont{and}
  \bibinfo{author}{\bibfnamefont{W.}~\bibnamefont{Vogel}},
  \bibinfo{journal}{Phys. Rev. Lett.} \textbf{\bibinfo{volume}{89}},
  \bibinfo{pages}{283601} (\bibinfo{year}{2002}).

\bibitem[{\citenamefont{Kiesel et~al.}()\citenamefont{Kiesel, Vogel, Parigi,
  Zavatta, and Bellini}}]{Kiesel08}
\bibinfo{author}{\bibfnamefont{T.}~\bibnamefont{Kiesel}},
  \bibinfo{author}{\bibfnamefont{W.}~\bibnamefont{Vogel}},
  \bibinfo{author}{\bibfnamefont{V.}~\bibnamefont{Parigi}},
  \bibinfo{author}{\bibfnamefont{A.}~\bibnamefont{Zavatta}}, \bibnamefont{and}
  \bibinfo{author}{\bibfnamefont{M.}~\bibnamefont{Bellini}},
  \eprint{arXiv:0804.1016v1}.

\bibitem[{\citenamefont{Korbicz et~al.}(2005)\citenamefont{Korbicz, Cirac,
  Wehr, and Lewenstein}}]{Korbicz05}
\bibinfo{author}{\bibfnamefont{J.~K.} \bibnamefont{Korbicz}},
  \bibinfo{author}{\bibfnamefont{J.~I.} \bibnamefont{Cirac}},
  \bibinfo{author}{\bibfnamefont{J.}~\bibnamefont{Wehr}}, \bibnamefont{and}
  \bibinfo{author}{\bibfnamefont{M.}~\bibnamefont{Lewenstein}},
  \bibinfo{journal}{Physical Review Letters} \textbf{\bibinfo{volume}{94}},
  \bibinfo{eid}{153601} (\bibinfo{year}{2005}).

\bibitem[{\citenamefont{Cahill and Glauber}(1969)}]{Cahill69}
\bibinfo{author}{\bibfnamefont{K.~E.} \bibnamefont{Cahill}} \bibnamefont{and}
  \bibinfo{author}{\bibfnamefont{R.~J.} \bibnamefont{Glauber}},
  \bibinfo{journal}{Phys. Rev.} \textbf{\bibinfo{volume}{177}},
  \bibinfo{pages}{1882} (\bibinfo{year}{1969}).

\bibitem[{\citenamefont{Hudson}(1974)}]{Hudson74}
\bibinfo{author}{\bibfnamefont{R.~L.} \bibnamefont{Hudson}},
  \bibinfo{journal}{Rep. Math. Phys.} \textbf{\bibinfo{volume}{6}},
  \bibinfo{pages}{249} (\bibinfo{year}{1974}).

\bibitem[{\citenamefont{Kenfack and \.{Z}yczkowski}(2004)}]{Kenfack04}
\bibinfo{author}{\bibfnamefont{A.}~\bibnamefont{Kenfack}} \bibnamefont{and}
  \bibinfo{author}{\bibfnamefont{K.}~\bibnamefont{\.{Z}yczkowski}},
  \bibinfo{journal}{J. Opt. B: Quantum Semiclass. Opt.}
  \textbf{\bibinfo{volume}{6}}, \bibinfo{pages}{396} (\bibinfo{year}{2004}).

\bibitem[{\citenamefont{Agarwal}(1981)}]{Agarwal81}
\bibinfo{author}{\bibfnamefont{G.~S.} \bibnamefont{Agarwal}},
  \bibinfo{journal}{Phys. Rev. A} \textbf{\bibinfo{volume}{24}},
  \bibinfo{pages}{2889} (\bibinfo{year}{1981}).

\bibitem[{\citenamefont{Hannay and Berry}(1980)}]{Hannay80}
\bibinfo{author}{\bibfnamefont{J.}~\bibnamefont{Hannay}} \bibnamefont{and}
  \bibinfo{author}{\bibfnamefont{M.~V.} \bibnamefont{Berry}},
  \bibinfo{journal}{Physica D} \textbf{\bibinfo{volume}{1}},
  \bibinfo{pages}{267} (\bibinfo{year}{1980}).

\bibitem[{\citenamefont{Dowling et~al.}(1994)\citenamefont{Dowling, Agarwal,
  and Schleich}}]{Dowling94}
\bibinfo{author}{\bibfnamefont{J.~P.} \bibnamefont{Dowling}},
  \bibinfo{author}{\bibfnamefont{G.~S.} \bibnamefont{Agarwal}},
  \bibnamefont{and} \bibinfo{author}{\bibfnamefont{W.~P.}
  \bibnamefont{Schleich}}, \bibinfo{journal}{Phys. Rev. A}
  \textbf{\bibinfo{volume}{49}}, \bibinfo{pages}{4101} (\bibinfo{year}{1994}).

\bibitem[{\citenamefont{Rivas and de~Almeida}(1999)}]{Rivas99}
\bibinfo{author}{\bibfnamefont{A.~M.~F.} \bibnamefont{Rivas}} \bibnamefont{and}
  \bibinfo{author}{\bibfnamefont{A.~M.~O.} \bibnamefont{de~Almeida}},
  \bibinfo{journal}{Ann. Phys. (N.Y.)} \textbf{\bibinfo{volume}{276}},
  \bibinfo{pages}{223} (\bibinfo{year}{1999}).

\bibitem[{\citenamefont{Bianucci et~al.}(2002)\citenamefont{Bianucci, Miquel,
  Paz, and Saraceno}}]{Bianucci02}
\bibinfo{author}{\bibfnamefont{P.}~\bibnamefont{Bianucci}},
  \bibinfo{author}{\bibfnamefont{C.}~\bibnamefont{Miquel}},
  \bibinfo{author}{\bibfnamefont{J.~P.} \bibnamefont{Paz}}, \bibnamefont{and}
  \bibinfo{author}{\bibfnamefont{M.}~\bibnamefont{Saraceno}},
  \bibinfo{journal}{Phys. Lett. A} \textbf{\bibinfo{volume}{297}},
  \bibinfo{pages}{353} (\bibinfo{year}{2002}).

\bibitem[{\citenamefont{Cormick et~al.}(2006)\citenamefont{Cormick, Galv{\~a}o,
  Gottesman, Paz, and Pittenger}}]{Cormick06}
\bibinfo{author}{\bibfnamefont{C.}~\bibnamefont{Cormick}},
  \bibinfo{author}{\bibfnamefont{E.~F.} \bibnamefont{Galv{\~a}o}},
  \bibinfo{author}{\bibfnamefont{D.}~\bibnamefont{Gottesman}},
  \bibinfo{author}{\bibfnamefont{J.~P.} \bibnamefont{Paz}}, \bibnamefont{and}
  \bibinfo{author}{\bibfnamefont{A.~O.} \bibnamefont{Pittenger}},
  \bibinfo{journal}{Phys. Rev. A} \textbf{\bibinfo{volume}{73}},
  \bibinfo{eid}{012301} (\bibinfo{year}{2006}).

\bibitem[{\citenamefont{Gross}(2006)}]{Gross06}
\bibinfo{author}{\bibfnamefont{D.}~\bibnamefont{Gross}}, \bibinfo{journal}{J.
  Math. Phys.} \textbf{\bibinfo{volume}{47}}, \bibinfo{eid}{122107}
  (\bibinfo{year}{2006}).

\bibitem[{\citenamefont{Galv{\~a}o}(2005)}]{Galvao05}
\bibinfo{author}{\bibfnamefont{E.~F.} \bibnamefont{Galv{\~a}o}},
  \bibinfo{journal}{Phys. Rev. A} \textbf{\bibinfo{volume}{71}},
  \bibinfo{eid}{042302} (\bibinfo{year}{2005}).

\bibitem[{\citenamefont{Arecchi et~al.}(1972)\citenamefont{Arecchi, Courtens,
  Gilmore, and Thomas}}]{Arecchi72}
\bibinfo{author}{\bibfnamefont{F.~T.} \bibnamefont{Arecchi}},
  \bibinfo{author}{\bibfnamefont{E.}~\bibnamefont{Courtens}},
  \bibinfo{author}{\bibfnamefont{R.}~\bibnamefont{Gilmore}}, \bibnamefont{and}
  \bibinfo{author}{\bibfnamefont{H.}~\bibnamefont{Thomas}},
  \bibinfo{journal}{Phys. Rev. A} \textbf{\bibinfo{volume}{6}},
  \bibinfo{pages}{2211} (\bibinfo{year}{1972}).

\bibitem[{\citenamefont{Lewenstein et~al.}(2000)\citenamefont{Lewenstein,
  Bruss, Cirac, Kraus, Kus, Samsonowicz, Sanpera, and Tarrach}}]{Lewenstein00}
\bibinfo{author}{\bibfnamefont{M.}~\bibnamefont{Lewenstein}},
  \bibinfo{author}{\bibfnamefont{D.}~\bibnamefont{Bruss}},
  \bibinfo{author}{\bibfnamefont{J.~I.} \bibnamefont{Cirac}},
  \bibinfo{author}{\bibfnamefont{B.}~\bibnamefont{Kraus}},
  \bibinfo{author}{\bibfnamefont{M.}~\bibnamefont{Kus}},
  \bibinfo{author}{\bibfnamefont{J.}~\bibnamefont{Samsonowicz}},
  \bibinfo{author}{\bibfnamefont{A.}~\bibnamefont{Sanpera}}, \bibnamefont{and}
  \bibinfo{author}{\bibfnamefont{R.}~\bibnamefont{Tarrach}},
  \bibinfo{journal}{J. Mod. Optics} \textbf{\bibinfo{volume}{47}},
  \bibinfo{pages}{2841} (\bibinfo{year}{2000}).

\bibitem[{\citenamefont{Abramowitz and Stegun}(1965)}]{Abramowitz65}
\bibinfo{editor}{\bibfnamefont{E.}~\bibnamefont{Abramowitz}} \bibnamefont{and}
  \bibinfo{editor}{\bibfnamefont{I.~A.} \bibnamefont{Stegun}}, eds.,
  \emph{\bibinfo{title}{Handbook of Mathematical Functions}}
  (\bibinfo{publisher}{Dover}, \bibinfo{year}{1965}).

\bibitem[{\citenamefont{Cormen et~al.}(2001)\citenamefont{Cormen, Leiserson,
  Rivest, and Stein}}]{Cormen01}
\bibinfo{author}{\bibfnamefont{T.~H.} \bibnamefont{Cormen}},
  \bibinfo{author}{\bibfnamefont{C.~E.} \bibnamefont{Leiserson}},
  \bibinfo{author}{\bibfnamefont{R.~L.} \bibnamefont{Rivest}},
  \bibnamefont{and} \bibinfo{author}{\bibfnamefont{C.}~\bibnamefont{Stein}},
  \emph{\bibinfo{title}{Introduction to Algorithms}}, vol. \bibinfo{volume}{2nd
  edition} (\bibinfo{publisher}{MIT Press and McGraw-Hill},
  \bibinfo{year}{2001}).

\bibitem[{\citenamefont{Peres}(1993)}]{Peres93}
\bibinfo{author}{\bibfnamefont{A.}~\bibnamefont{Peres}},
  \emph{\bibinfo{title}{{Quantum Theory: Concepts and Methods}}}
  (\bibinfo{publisher}{Kluwer Academic Publishers},
  \bibinfo{address}{Dordrecht}, \bibinfo{year}{1993}).

\bibitem[{\citenamefont{Horodecki et~al.}(1996)\citenamefont{Horodecki,
  Horodecki, and Horodecki}}]{Horodecki96}
\bibinfo{author}{\bibfnamefont{M.}~\bibnamefont{Horodecki}},
  \bibinfo{author}{\bibfnamefont{P.}~\bibnamefont{Horodecki}},
  \bibnamefont{and}
  \bibinfo{author}{\bibfnamefont{R.}~\bibnamefont{Horodecki}},
  \bibinfo{journal}{Phys. Lett. A} \textbf{\bibinfo{volume}{232}},
  \bibinfo{pages}{1} (\bibinfo{year}{1996}).

\end{thebibliography}

\end{document}